\documentclass[aps,prl,twocolumn,showpacs,superscriptaddress,groupedaddress]{ 
revtex4-1}  
\usepackage{graphicx}  
\usepackage{dcolumn}   
\usepackage{bm}        
\usepackage{amssymb}   
\usepackage{amsmath} 
\usepackage{mathrsfs}  
\usepackage{graphicx,color} 
   
\begin{document}

\title{Detecting topological sectors in continuum Yang-Mills theory and the fate of BRST symmetry}

\author{L.~E.~Oxman and G.~C.~Santos-Rosa}
 
\affiliation{
Instituto de F\'\i sica, Universidade Federal Fluminense,
24210-346 Niter\'oi - RJ, Brasil.}

\date{\today}

\begin{abstract}

In this work, motivated by Laplacian type center gauges in the lattice, designed to avoid the Gribov problem, we introduce a new family of gauge fixings for pure Yang-Mills theories in the continuum. This procedure separates the partition function into partial contributions associated with different sectors, containing center vortices and correlated monopoles. 
We show that, on each sector, the gauge fixed path-integral displays a BRST symmetry, however, it cannot be globally extended due to sector dependent boundary conditions on the ghost fields. These are nice features as they would permit to discuss the independence of the partial contributions on gauge parameters,, while opening a window for the space of quantum states to be different from the perturbative one, which would be implied if topological configurations were removed.

\end{abstract}

\pacs{include}
\maketitle

\section{Introduction}

The quantum behaviour of a theory is encoded in the path-integral measure. In
the case of Yang-Mills (YM) theory the main problem is to implement the sum over 
physically inequivalent configurations. At the perturbative level, this can be 
done by following the Faddeev-Popov procedure which introduces an 
identity based on a gauge fixing condition for the gauge field, enforced 
by Lagrange multiplier auxiliary fields. This permits to factor out the group 
volumes at different spacetime points, and get an action that 
contains ghost fields. Upon gauge fixing, the obtained action displays a BRST 
symmetry that plays a key role to construct the (perturbative) space of physical 
states. On the other hand, at the nonperturbative level, usual gauge fixing conditions
such as the Landau, Coulomb and maximally Abelian gauges lead to 
the Gribov problem. That is, upon gauge fixing there are still gauge copies. To avoid infinitesimal copies, Gribov proposed the path-integral restriction to the region where the Faddeev Popov operator is positive definite \cite{gribov}, and the idea that this could modify the infrared gluon propagator, leading to gluon confinement. His procedures were implemented through a modified action \cite{zwanziger1}, and gluon confinement has been extensivelly discussed through different breaking scenarios for BRST symmetries (see \cite{schaden}, \cite{silvio-soft}, and refs. therein). In this respect, a BRST symmetry that takes the horizon function into account was recently proposed \cite{silvio-hor}.  
 
In the lattice, examples of gauges that are designed to avoid the Gribov problem were given in refs. \cite{VW}, \cite{deforcrand}, \cite{faber} (for a review, see ref. \cite{greensite-livro}). 
In ref. \cite{faber}, the lowest $d=N^2-1$ eigenfunctions 
$\zeta_1, \dots, \zeta_d$ are computed, and an algorithm is given that associates 
them with a local Lie basis  
$ST_1S^{-1}, \dots, ST_d S^{-1}$, $S\in SU(N)$. Finally, the gauge is fixed by transforming this basis to a given reference. This gauge, known as the direct Laplacian center gauge, properly detects center vortices. These are topological degrees of freedom which, together with correlated monopoles, seem to be essential to understand quark confinement due to a Wilson loop area law, and obtain a confining potential with the right properties (see \cite{greensite-livro} and refs. therein). These global aspects are the essence of dual superconductivity, which is based on topological magnetic degrees of freedom that condense \cite{thooft}. This is also the case in the compact version of QED, which is confining \cite{polyakov}.  From this perspective, it would be desirable to introduce a gauge fixing in the continuum that is appropriate when applied to genuine gauge field variables in topological sectors. 

Indeed, the Gribov problem seems to be originated from the fact that usual gauge fixing conditions in the continuum take values on the Lie algebra \cite{singer}, the tangent space to the Lie group.  The problem is manifested when such gauge conditions are applied to configurations whose nature depends on the global aspects of the group. Note that topological configurations lie on the Gribov horizon \cite{greensite-zwanziger}, \cite{maas}, where the Faddeev-Popov operator vanishes. 
Then,  it is appropriate to turn our attention to gauge conditions in the continuum that take values on group elements.
In this letter, we shall initially look for a procedure that uniquely associates the gauge field $\mathscr{A}_\mu$ with a field $T(S)$, in a covariant way. Here $T(G)$, $S\in G$, stands for a representation of the gauge group. Then, for this gauge field, we shall define the gauge condition requiring $T(S) = T(S_0)$, where $S_0$ is a prescribed map. For convenience, we define the theory on the Euclidean spacetime. 

For $SU(N)$ Yang-Mills with a gauge fixing based on the adjoint representation, we propose that the role of the variable $S \in SU(N)$ be played by 
a non Abelian phase in a generalized polar decomposition of a tuple $(\zeta_1, \zeta_2,\dots)$, where each scalar field carries an adjoint 
representation of the gauge group. To produce a strong correlation between $\mathscr{A}_\mu$ and these phases, 
the scalars are required to be classical solutions in a model with $SU(N) \to Z(N)$ SSB. In this manner, the tuples will locally tend to be as close as possible to a point on the manifold of vacua $Ad(SU(N))$, that can be used to implement the gauge fixing. It is important to underline that the SSB pattern does not mean we will consider the dynamics of a Yang-Mills-Higgs model. Our path integral weight does not contain the Higgs factor, and additional Grassmann fields $c_I$, $\bar{c}_I$ will also be introduced
to keep the dynamics to be that of pure Yang-Mills.

The important point is that it is not possible to use the same reference $S_0$ for all the variables $\mathscr{A}_\mu$ that must be path-integrated. 
Although $\mathscr{A}_\mu$ is smooth and well-defined everywhere, the associated $S$ could contain defects and, in that case, it cannot be used to perform gauge transformations. Instead, the variables have to be splitted into infinitely many sectors $\mathscr{V}(S_0)$.
One of them is given by those fields $\mathscr{A}_\mu$ whose associated non Abelian phase $T(S)$ contains no defects. 
In this sector, $S_0$ can be simply chosen as the identity map $S_0 \equiv I$. In other sectors, $S_0$ could contain center vortex and monopole-like  defects, as well as correlated mixtures. 

Here, we shall not attempt the definition of a measure that contemplates all the different sectors at once, which is a hard problem. 
We shall pursue instead a gauge fixing on each sector $\mathscr{V}(S_0)$.  
Similarly to the lattice, where center vortex and monopole removal implies a Wilson loop perimeter law, while their inclusion leads to the area law, we shall analyze the effect of ``turning on and off" the defects on the definition of a BRST transformation for the full partition function. If all sectors labelled by mappings $S_0$ that contain defects were removed from the path-integral, we will show there is a well-defined BRST symmetry. On the other hand, the consideration of the complete theory will interfere with the definition of a BRST charge operator and the ensuing analysis of physical states.

\section{Gauge fixing procedure}

The search for the mapping $\mathscr{A}_\mu \to T(S) $ is guided by the property of covariance. If $\mathscr{A}_\mu$ is associated with $T(S)$, then the gauge transformed field $\mathscr{A}_\mu^U = U \mathscr{A}_\mu U^{-1} + (i/g)\,  U\partial_\mu U^{-1}$, where $U : R^4 \to G$ is a regular non Abelian phase, must be associated with $T(US)$. If this is a well-defined function that maps a given $\mathscr{A}_\mu$ into a unique field $T(S)$, then no Gribov copies would be present. 
In this respect, consider any regular gauge transformation $\mathscr{A}^U_\mu$ and suppose it also satisfies the gauge condition. Covariance would give $T(US_0) = T(U)T(S_0) = T(S_0) $, that is, 
$T(U)$ is the identity mapping. From now on, we will take $SU(N)$ Yang-Mills theory with a gauge fixing based on the adjoint representation. The equation for copies implies that $U$ is 
in the center $Z(N)$. As this is a discrete group, $U$ must be $x$-independent so that $\mathscr{A}_\mu^U = \mathscr{A}_\mu$, i.e., neither infinitesimal nor finite Gribov copies would be present. 

To define the mapping, we initially relate the non Abelian gauge field 
$\mathscr{A}_\mu$ with a set of hermitian fields $\zeta_I \in \mathfrak{su}(N)$, 
where $I$ is a flavour index. These fields minimize an action $S_H$, in the presence of 
$\mathscr{A}_\mu$,
\begin{equation}
\frac{\delta S_H}{\delta \psi_I}\bigg|_{\zeta_I} = 0 \makebox[.3in]{,}
S_H = \int d^4 x\, \frac{1}{2}\,\langle \mathscr{D}_\mu \psi_I \rangle^2 + V_{H} 
\;,
\label{sol}
\end{equation}
$\mathscr{D}_{\mu} = \partial_{\mu} -ig[\mathscr{A}_{\mu}, \;]$, which is gauge invariant
under 
\[\mathscr{A}_\mu \to \mathscr{A}_\mu^U \makebox[.3in]{,} \psi_I \to \psi_I^U = U \psi_I U^{-1}\;, \]
and the potential $V_H \geq 0$ is constructed with an $SU(N) \to Z(N)$ SSB pattern.
Here, we used the positive definite 
internal product in the Lie algebra,
\begin{equation}
\label{metric} 
\langle X, Y \rangle = Tr(Ad(X)Ad(Y)) \;,
\end{equation}
with $Ad(X)$ given by a map of $X \in \mathfrak{su}(N)$ into the adjoint representation. As  
a shorthand notation, we use $\langle X  \rangle^2=\langle X, X \rangle$.  
The points in the manifold ${\cal M}$ of absolute minima of the potential shall 
be denoted by the tuple $ (v\, \phi_1, v\,\phi_2, \dots )$.
The parameter $v$ has mass dimension one, while the Lie algebra elements 
$\phi_I$ are dimensionless. At $\psi_I = v\,\phi_I$, it is verified 
$\frac{\delta V_H}{\delta \psi_I}=0 $, and $V_H = 0$.
As the field configurations $\mathscr{A}_\mu$ that contribute to the Yang-Mills partition function 
$Z_{YM}$ (locally) tend to a pure gauge at infinity, the minimization of $S_H$ requires the boundary 
conditions, 
\begin{equation}
\zeta_I \to v\, n_I \makebox[.5in]{,} \mathscr{D}_\mu n_I \to 0 \;,
\label{asy}
\end{equation}
where  $(v\,n_1 , v\, n_2, \dots )\in {\cal M}$. 
 
Next, given a tuple $(\psi_1, \psi_2, \dots)$, we would like to define 
``modulus'' $q_I$ and ``phase'' $S$ variables, by means of the ``polar''  decomposition $\psi_I = Sq_IS^{-1}$,
together with a set of conditions $f_A(q_1,q_2,\dots)=0$, $A=1,\dots, d=N^2-1$. In our case, it is 
natural to define $S$ from $\phi_I = S u_I S^{-1}$, where $(v\, u_1, 
v\, u_2,...)\in {\mathcal{M}}$  is an $x$-independent reference point, and
$(v\, \phi_1, v\,\phi_2,...) \in {\mathcal{M}}$ is the point closer to $(\psi_1, \psi_2,...)$, obtained by minimizing  $\sum_I \langle \psi_I - v\, \phi_I\rangle^{2}$. This phase and modulus concept correspond to having tuples
$(v\, \phi_1, v\, \phi_2,...) \in {\mathcal{M}}$ and $(q_1, q_2,...)$ that are aligned as much as possible (in the mean) with respect to    $(\psi_1, \psi_2, \dots)$ and  $(u_1, u_2, \dots)$, respectively,
\begin{equation}
\sum_I [\psi_I, \phi_I] = 0
\makebox[.3in]{,}
\sum_I [q_I, u_I] = 0 \;.
\label{fes} 
\end{equation}
Note that the first equation, obtained by minimizing the average distance, together with $\phi_I = S u_I S^{-1}$, implies the second. 
In this case, the functions $f_A$ are obtained by projecting eq. (\ref{fes}) with a Lie basis.

Now, given a field $\mathscr{A}_\mu$, the solution $(\zeta_1,\zeta_2, \dots)$ that satisfies the boundary conditions (\ref{asy}) is expected to be unique and correspond to a stable absolute minimum of the action. In addition, because of the $SU(N) \to Z(N)$ SSB pattern, 
the polar angle $S$ corresponding to $(\zeta_1,\zeta_2, \dots)$ is unique up to a global center transformation (almost everywhere).  In effect, if $S$ and $S'$ represent the point on the manifold of vacua $\mathcal{M}$ closest to $(\zeta_1,\zeta_2, \dots)$, we can imply $S u_I S^{-1} = S' u_I {S'}^{-1}$. In other words, the adjoint transformation based on $S^{-1} S'$ leaves the point $(v\, u_1, v\, u_2, \dots) \in \mathcal{M}$ invariant, and as the invariance group of a point in $\mathcal{M}$ is $Z(N)$, we get $S' = z\, S$, with $z\in Z(N)$.
Finally, as the center is a discrete group, the factor $z$ must be global. Then, the sequence,
\begin{equation}
\mathscr{A}_\mu \to \zeta_I \to v\, \phi_I \to Ad(S)  \;,  
\label{seq}
\end{equation}
gives a well-defined mapping from $\mathscr{A}_\mu \to Ad(S)$, where $Ad$ stands for the adjoint representation.
In addition, due to covariance of the field equations and the polar decomposition, as well as the group invariance of the metric, the solution for the gauge transformed field $\mathscr{A}_{\mu}^{{U}} $ is given by 
$U\zeta_IU^{-1} $, which corresponds to the polar non Abelian phase $Ad(US)$. Therefore, when fixing the gauge $Ad(S)=Ad(S_0)$, in the $\mathscr{V}(S_0)$ sector,
the conditions discussed above, needed for the absence of Gribov copies, are in principle satisfied. 

The following comments are in order. To implement the required SSB pattern for the classical scalar fields, the minimum number of flavours is $N$. A natural model based on $N^2-1$ flavours, with global flavour symmetry $Ad(SU(N))$, was introduced in ref. \cite{oxman}. In this case, renaming $I\to A=1, \dots, d=N^2-1$, the potential is,
\begin{eqnarray}
&& V_H(\psi)=\frac{\mu^{2}}{2}\, \langle \psi_A, \psi_A \rangle + \frac{\kappa}{3}\, f_{ABC} \langle \psi_A, \psi_B \wedge \psi_C \rangle \nonumber \\
&&  + 
\frac{\lambda}{4}\, \langle \psi_A \wedge \psi_B, \psi_A \wedge \psi_B \rangle  \;,
\end{eqnarray}
where $\psi_A \wedge \psi_B = -i [\psi_A, \psi_B]$, and $f_{ABC}$ are $\mathfrak{su}(N)$ structure constants.  When $m^2 < \frac{2}{9}\frac{\kappa^{2}}{\lambda}$, it induces $SU(N) \to Z(N)$ SSB. For $\kappa < 0$, the manifold $\mathcal{M}$ is  characterized by  $v =-\frac{\kappa}{2\lambda}\pm [ (\frac{\kappa}{2\lambda})^{2}-\frac{\mu^{2}}{\lambda}]^{\frac{1}{2} } $ and $\phi_A = S u_A S^{-1}$, where the reference point $u_A$ is an $x$-independent Lie basis, $[u_A, u_B] = if_{ABC}\, u_C$.
 
In this case, expanding $\psi_A =\Psi|_{AB}\, u_B$, $q_A =Q|_{AB}\, u_B$, the polar decomposition $\psi_A = S q_A S^{-1}$  of the tuple $(\psi_1, \dots, \psi_d)$ corresponds to $\Psi = Q\, R(S)$, where $R(S)$ gives the map of $S\in SU(N)$ into the adjoint representation, $Su_AS^{-1}= R|_{AB}\, u_B$. In addition, projecting the second equation in (\ref{fes}), the modulus concept becomes, 
\begin{equation}
f_A = \sum_B \langle u_A, q_B \wedge u_B\rangle = 0 \;, 
\end{equation}
that is, ${\rm tr}\, (QM_A) =0$, where $M_A$ are adjoint generators, $M_A|_{BC}=-i f_{ABC}$. This together with the property
$g_\psi = v^2\, Q\, g_u\, Q^T$, where $g_\psi|_{AB} = \langle \psi_A, \psi_B \rangle$, 
$g_u|_{AB} = \langle u_A, u_B \rangle = {\rm tr}\, (M_A M_B)$, determine $Q$. It is worth emphasizing that the metric $g_u$ is $x$-independent, on the other hand, $g_\psi$ is $x$-dependent.
 
For example, when $N=2$, $\Psi$ is decomposed in terms of $Q$, orthogonal to $M_A$,
and $R(S) \in Ad(SU(2))=SO(3)$. As the matrices $M_A$, $A=1,2,3$ form a basis for the $3\times 3$ antisymmetric matrices, $Q$ must be symmetric, and our procedure gives the usual polar decomposition of a $3\times 3$ real matrix $\Psi$.

We emphasize that in ref. \cite{oxman}, a similar SSB pattern was used to construct a dual superconductor model in the Higgs phase, where smooth center vortices represent confining strings between quarks, implementing $N$-ality. Here, 
the classical scalars only provide a means to fix the gauge. In the path integral, they will be complemented with Grassman fields, so as to keep the pure Yang-Mills dynamics unchanged. 

\section{Gauge fixed path integral in a general sector $\mathscr{V}(S_0)$} 

In principle, we are interested in the Yang-Mills partition function $Z_{YM} =  \int [D\mathscr{A}_\mu] \; e^{-S_{YM}(\mathscr{A})}$. As discussed in the previous sections, the sequence \eqref{seq} will assign the variable $\mathscr{A}_\mu$ with a field $Ad(S)$ that could contain defects. Then, to implement our gauge fixing, this path integral must be separated as an infinite sum of partition functions, 
\begin{eqnarray}
\label{ymf}
&& Z_{YM}^{(S_0)}=  \int_{\mathscr{V} (S_0)} [D\mathscr{A}_\mu] \; e^{-S_{YM}(\mathscr{A})} \;, \\
&& S_{YM}(\mathscr{A}) =\int d^4 x\, \frac{1}{4}\langle \mathscr{F}_{\mu\nu}\rangle^2 \makebox[.3in]{,}\mathscr{F}_{\mu\nu} = \frac{i}{g}\,[\mathscr{D}_{\mu},\mathscr{D}_{\nu} ]
\;, \nonumber
\end{eqnarray}
computed over sectors $\mathscr{V}(S_0)$ where $\mathscr{A}_\mu \to Ad(S)$, with $S=U S_0$ and $U$ being a regular map. Of course, different labels $S_0, S'_0$ must be such that there is no regular $U$ connecting them in the form $S'_0 = U S_0$.

The solution $\zeta_I$ to eq. (\ref{sol}), in the presence of $\mathscr{A}_\mu$, 
can be introduced by following similar techniques to those given in the context of stochastic field equations (see for example refs. \cite{Zinn-justin}, \cite{DZ}).  Here, in order to keep the path integral unchanged, we introduce the identity,   
\begin{equation}
\label{deltapsi}
1 =\int [D\psi_I]\, \prod_{I}\delta(\psi_{I} - \zeta_{I}) \;.
\end{equation}
The delta functional can be rewritten as,
\begin{equation}
 \prod_{I}\delta(\psi_{I} - \zeta_{I}) = \det \left( \frac{\delta^{2}S_H}{\delta 
\psi_{J} \delta \psi_{I}}\right)\, \prod_{I} \delta\left(\frac{\delta 
S_H}{\delta \psi_{I}}\right)  \;,
 \label{deltaf}
\end{equation}
where the first and second derivatives are defined by the expansion,
\begin{eqnarray}
S_H(\psi + \delta\psi)= S_H(\psi) + \langle \delta\psi_I, \frac{\delta 
S_H}{\delta \psi_{I}} \rangle \nonumber \\+ \frac{1}{2} \langle \delta\psi_I, 
\frac{\delta^{2}S_H}{\delta \psi_{J} \delta \psi_{I}} \delta\psi_J \rangle + 
{\cal O}(\delta \psi^3) \;. 
\end{eqnarray}
The boundary conditions and the SSB pattern, that are important to discuss the absence of Gribov copies, 
here reappear as follows. As they would lead to a solution $\zeta_I$ that is unique, and corresponds to a stable absolute minimum of the action, eq. (\ref{deltaf}) is expected to be a well-defined representation of the delta 
functional. In particular, the operator $\frac{\delta^{2}S_H}{\delta \psi_{J} \delta \psi_{I}}  $ 
would be associated with a positive definite quadratic form. Then, the 
partition function (\ref{ymf}) is,
\begin{equation}
Z_{YM}^{(S_0)}= \int_{\mathscr{V} (S_0)} [D\mathscr{A}_\mu][D\psi_I][D\xi_I] [D\bar{\omega}_I][D\omega_I]    \; 
e^{-\mathscr{S}_f} \;,
\label{pf}
\end{equation}
where $\mathscr{S}_f$ is the Yang-Mills action extended by the addition of auxiliary and ghost fields, 
\begin{eqnarray}
\label{ymfsh}
\lefteqn{\mathscr{S}_f  =S_{YM}(\mathscr{A}) } \nonumber \\ && + \int d^{4}x 
\; \left( \langle \mathscr{D}_\mu \bar{\omega}_{I}, \mathscr{D}_\mu \omega_{I} 
\rangle + \langle \bar{\omega}_{I}, \frac{\delta^{2}V}{\delta \psi_{J} \delta 
\psi_{I}}\; \omega_{J} \rangle \right) \nonumber \\ && +  \int d^{4}x \; \left( 
\langle \mathscr{D}_\mu \xi_{I}, \mathscr{D}_\mu \psi_{I} \rangle + \langle 
\xi_{I}, \frac{\delta V}{\delta \psi_I} \rangle \right) \; . 
\end{eqnarray}
The fields $\xi_I$ (resp. $\bar{\omega}_I$, $\omega_{I}$) are adjoint bosonic 
(resp. fermionic) fields. 

To obtain the gauge fixed $Z_{YM}^{(S_0)}$, we still need to change variables from the tuple $\psi_I$ to a path integral over the variables in a polar decomposition, and factor out the group volume $[DU]$ over regular transformations. 
The path-integral over the tuples in eq.  \eqref{deltapsi} only receive a contribution when $\psi_I$ is the solution $\zeta_I$ to the classical equations of motion. Then, in the $\mathscr{V}(S_0)$ sector, the relevant tuples are of the form $\psi_I = S q_I S^{-1} $, with $f_A(q_I)=0$, $S=US_0 $, and $U$ regular. 

In order to change in the path integral to modulus and phase variables, let us momentarily forget about spacetime. Considering a single tuple $(\psi_1, \psi_2, \dots)$, and a function $F(\psi_1, \psi_2, \dots)$, we have \cite{Adler}, 
\begin{eqnarray}
\label{muddevar}
\int d\psi_I\, F(\psi_I) &=& \int dU dq_I \; \delta (f_A(q) ) \det \left[J(q)\right]\,  
F(UqU^{-1})\;, \nonumber \\
 J(\psi)|_{AB}&=& \frac{\partial f_A (\psi + [\varepsilon,\psi])}{\partial 
\varepsilon_B}\bigg|_{\varepsilon=0} \;. 
\end{eqnarray}
The notation $UqU^{-1}$ means $(Uq_1U^{-1}, Uq_2U^{-1}, \dots)$.
In the flavour symmetric model, $f_A={\rm tr} (QM_A)$ leads to
$J_{AB}(q)= {\rm tr}\, (M_A\,Q\,M_B)$. For $N=2$, $J(q)= {\rm tr}(Q)\, \delta_{AB} - Q_{AB}$, and  
\begin{equation} 
\label{jacsu2}
 \det \left[J(q)\right] = \prod_{i<j} (\lambda_i + \lambda_j) \;,   
\end{equation}
where $\lambda_i$ are the eigenvalues of $Q$. In addition, in this case $Q$ is symmetric, so
it can be diagonalized by a matrix $R\in SO(3)$. The change of variables gives 
$\int dq = \int dR d\lambda \, \prod_{i<j} (\lambda_i - \lambda_j) $ where the new Jacobian was  
borrowed from the orthogonal ensemble, in random matrix theory. As expected, the product of both Jacobians 
gives that corresponding to the ensemble of real $3\times 3$ matrices $\Psi$, $\prod_{i<j} (\lambda_i^2 - \lambda_j^2) $ \cite{matale}.

Equation (\ref{muddevar}) is based on the premise that the variables $(\psi_1, \psi_2, \dots )$ can be rotated to satisfy $f_A = 0$. Then, if $T(S)$ contains no defects, we can take the reference as $S_0 = I$,
use a direct generalization of eq. (\ref{muddevar}) to path-integrals, and replace it into eq. (\ref{pf}). 
In a general sector characterized by $S_0$, we can use the following identity,
\[
1= \int [DU] \, \delta(f_A(S^{-1}\psi_I S)) \det(J(q))
\makebox[.3in]{,}
S=US_0\;,
\]
to obtain,
\begin{eqnarray}
\lefteqn{ \int [D\psi_I] \, F[\psi_I] =} \nonumber \\
&& =\int [DU] [Dq_I] \, \delta(f_{A}(q)) \det(J(q)) F[S q_{I} S^{-1}] \;.
\end{eqnarray}
This generalizes eq. \eqref{muddevar} to spacetime dependent tuples and a general situation where the corresponding $x$-dependent 
polar angle $S$ could contain defects.
Next, we can use this change of variables in eq. \eqref{pf}, followed by,
\begin{equation} 
 \mathscr{A}_{\mu} = A_{\mu}^{U} \makebox[.3in]{,} \xi_I = b_{I}^{U} 
  \makebox[.3in]{,} 
 \bar{\omega}_{I} = \bar{c}_{I}^{U}  \makebox[.3in]{,}  \omega_I = 
c_{I}^{U} \;. 
\label{on}
\end{equation}
Then, using $[D A^{U}_\mu]=[DA_\mu]$, $[D\bar{c}^{U}_I] = [D\bar{c}_I]$, $[Dc^{U}_I] = [Dc_I]$ $[Db^{U}_I] = [Db_I]$, and the gauge invariance of $\mathscr{S}_f$, we can factor out the group volumes ${\cal N} = \int [DU]$. In this way we obtain the gauge fixed partition function, 
\begin{eqnarray}
Z_{YM}^{(S_0)}&=& {\cal N} \int [DA_\mu][Dq_I][Db_I][D\bar{c}_I][Dc_I]\times \nonumber \\ 
& \times &  \, \delta 
[f_A(q)]
\det [J(q)] \,e^{-S_f }  \;,
\label{funcgeri}
\end{eqnarray}
\begin{eqnarray}
\label{ymfgf}
\lefteqn{S_f =  S_{YM}(A) \, +} \nonumber \\  
&& + \int d^{4}x \; \left( \langle D_\mu \bar{c}_{I}, D_\mu c_{I} \rangle + 
\left\langle \bar{c}_{I}, \frac{\delta^{2}V}{\delta \psi_{J} \delta 
\psi_{I}}\Big|_{\psi_{I}= q_I^{S_0}}\; c_{J} \right\rangle \right)  \nonumber \\ 
&& + \int d^{4}x \; \left( \left\langle D_\mu b_{I}, D_\mu (q_I^{S_0}) \right\rangle + \left\langle 
b_{I}, \frac{\delta V}{\delta \psi_I}\Big|_{\psi_I = q_I^{S_0} } \right\rangle \right) 
\end{eqnarray}
\begin{eqnarray}
&&q_I^{S_0} = S_0 q_I S_{0}^{-1}  \makebox[.5in]{,} D_\mu= \partial_{\mu} - ig[ A_{\mu}\, , ~]  \;.
\end{eqnarray}
The presence of $S_0$ occurs as in general we cannot use the full map $S=US_0$ to perform gauge transformations.

\section{BRST in a general sector $\mathscr{V}(S_0)$} 

The action $\mathscr{S}_f$ in eq. (\ref{ymfsh}), where the gauge has not 
been fixed yet, possesses the following nilpotent symmetry,
\begin{eqnarray}
\label{brstant}
s\mathscr{A}_{\mu} = 0 \makebox[.2in]{,} s\psi_I = \omega_I \makebox[.2in]{,} 
s\xi_I = 0 \makebox[.2in]{,} s\bar{\omega}_I = -\xi_I \makebox[.2in]{,} 
s\omega_I = 0 \;. \nonumber \\
\end{eqnarray} 
Proceeding in a similar way to refs. \cite{tupper}, \cite{tupper2}, this can be used to 
analyze the BRST symmetry after the gauge fixing. 
Here, this will be done separately on each  sector generated by our procedure and then, in the conclusions,  we shall discuss the situation when the complete theory is considered.  As usual, using,
\begin{equation}
s\mathscr{A}_\mu = 0 = s(U A_{\mu} U^{\dagger} + \frac{i}{g} U \partial_\mu 
U^{\dagger}) \;,
\end{equation}
and the Faddeev-Popov ghost definition $c = U^{\dagger} s U$,
the BRST transformation of $A_\mu$ is,
\begin{equation}
\label{brsta}
s\, A_{\mu}=\frac{i}{g} D_{\mu}c  \;.
\end{equation}
Similarly, using the other equations in (\ref{on}), we get the BRST transformations,
\begin{eqnarray}
\label{brstq}
s\, q_I^{S_0} &=& [q_I^{S_0}, c ] + c_I 
, \\ \label{brstb} s\, b_I &=& [b_I, c], \\ 
\label{brstcb} s\, \bar{c}_I &=& -\{\bar{c}_I, c\} -b_I, \\ \label{brstc} 
s\, c_I &=& -\{c_I, c\} 
\;,
\end{eqnarray} 
which  together with $s\, S_0 =0$ and eq. (\ref{brsta}) constitute the nilpotent BRST symmetry of  $S_f $ in eq.  \eqref{ymfgf}. Indeed, we can write, 
\begin{equation}
 S_f = S_{YM} - \int d^{4}x\, s \left\langle \bar{c}_I, \frac{\delta S_H}{\delta \psi_{I}}{\big{|}}_{\psi_I=q_I^{S_0}} \right\rangle \;.
\end{equation}

Now, to determine the BRST symmetry of $Z_{YM}^{(S_0)}$, the gauge fixed partition function in $\mathscr{V}(S_0)$, we initially note that all transformations are combinations of infinitesimal rotations and translations. Then, the different measures $[DA_\mu], [Dq_I], \dots$ are invariant, and we are left with the problem of analyzing the effect on BRST of the remaining factor in eq. (\ref{funcgeri}), which is necessary for the variable $q_I$ to be a pure modulus.

As it arises when factoring the group volume, it is natural to introduce the representation,
\begin{eqnarray}
\lefteqn{\delta [f_A(q)] \; \det [J(q)] = \int [Db][D\bar{c}][Dc]} \nonumber \\
&& \times \; \exp{- \int d^{4}x\,  \left(\left\langle b, [u_I^{S_0}, q_I^{S_0}] \right\rangle + \left\langle  \bar{c} , [u_I^{S_0}, [q_I^{S_0} , c]] \right\rangle \right) } \nonumber \\
\label{jac} 
\end{eqnarray}
where the Grassmann variables have been identified with the ghost (antighost) fields $c$ ($\bar{c}$) in (\ref{brsta}),
and then complete the BRST transformations (\ref{brsta})-(\ref{brstc}) with,
\begin{equation}
\label{brste} 
sb=0 \makebox[.5in]{,} s\bar{c}=-b \, \makebox[.5in]{,} sc = - \frac{1}{2} \{c,c\}\;.
\end{equation}
Here, $f_{A}(q)$ has been based on the pure modulus condition in eq. \eqref{fes}.
Finally, we note that the exponent in eq. \eqref{jac} changes when the complete $s$-transformation is performed, due to the action of  
$s$ on $q_I$. However, it is easy to see that, upon integration over the ghost fields, the right-hand side of eq. \eqref{jac} remains unchanged. 
In order to evidence this symmetry, we can include in the integrand of eq. \eqref{jac} a factor $\exp ({-\int  d^{4}x\, \langle \bar{c}, [u_I^{S_0} , c_I] \rangle})$, which can be expanded as $1 -\int  d^{4}x\, \langle \bar{c}, [u_I^{S_0} , c_I] \rangle) + \dots$. It is clear that, in the integral over the ghost and antighost fields, only the first term gives a contribution. That is, we can also write,
\[
\delta [f_A(q)] \det [J(q)] = \int [Db][D\bar{c}][Dc] \; e^{\int d^{4}x \, s  \left\langle \bar{c}, [u_I^{S_0} , q_I^{S_0}] \right\rangle } \;,
\label{jaco}
\]
and thus obtain,
\begin{eqnarray}
\label{genfunexp2}
&&Z_{YM}^{(S_0)}= {\cal N} \int_{\mathscr{V} (S_0)} [DA_\mu][Dq_I][Db_I] [D\bar{c}_I][Dc_I] [Db][D\bar{c}][Dc] \nonumber \\ 
&& \times   \, 
e^{-S_{YM} + \int d^{4}x \, s \left( \left\langle \bar{c}, [u_I^{S_0}, q_I^{S_0} ] \right\rangle + \left\langle \bar{c}_I, \frac{\delta S_H}{\delta \psi_{I}}{\big{|}}_{\psi_I=q_I^{S_0}} \right\rangle \right) } \;.
\end{eqnarray}
In the sector of gauge fields $\mathscr{V}(I)$, where we simply have $q_I^{S_0} = q_I $, 
this representation is all we need to show the BRST invariance of $Z_{YM}^{(I)}$\,. However, in the sectors $\mathscr{V}(S_0)$, when $S_0$ contains defects, we still have to analyze the BRST invariance of the boundary conditions in the path-integral for $Z_{YM}^{(S_0)}$ , needed for the fields to be well-defined everywhere. The importance of the 
BRST invariance of boundary conditions have been discussed in different contexts, such as the Casimir effect \cite{moss}.

For example, let us consider variables $\mathscr{A}_\mu$ in a sector that contains center vortices around a given set of closed worldsheets. That is, we take $S_0 = e^{i\chi 
\vec{\beta}|_p T_p}$, where $\vec{\beta}=2N \vec{w}$, $\vec{w}$ is a weight of the fundamental representation, and $T_p$, $p=1, \dots, N-1$ are Cartan generators of $\mathfrak{su}(N)$. A Lie basis is completed with the root vectors $E_\alpha$, associated with positive and negative roots $\vec{\alpha}$. The phase $\chi$, $\partial^2 \chi =0$, is multivalued when we go around the worldsheets (see refs. \cite{reinhardt-engelhardt}, \cite{konishi-spanu}). In this case, $S_0 T_p S_0^{-1} =T_p$ and $S_0 E_\alpha S_0^{-1} = \cos (\vec{\beta}\cdot \vec{\alpha})\chi \,  E_\alpha + \sin (\vec{\beta}\cdot \vec{\alpha})\chi \,  E_{-\alpha}$.  Then, to have well-defined fields $\psi_I$, we must impose boundary conditions, at the worldsheets,
\begin{equation}
\left. \left\langle E_\gamma, q_I^{S_0} \right\rangle \right|_w =0 \makebox[.5in]{,} \vec{\beta}\cdot \vec{\gamma}\neq 0\;.
\label{bc}
\end{equation}
This means that the off-diagonal components of $q_I$, that are charged with respect to the weight $\vec{\beta}$, must vanish at the worldsheets. 
Applying $s$ on $\left\langle E_\gamma, q_I^{S_0} \right\rangle$, using eq. (\ref{brstq}), and then evaluating at the worldsheets, the BRST invariance of the boundary conditions would be,
\begin{equation}
\left. \left\langle E_\gamma,  [ q_I^{S_0}, c ]\right\rangle \right|_w +  \left\langle E_\gamma, c_I \right\rangle|_w   = 0  
\makebox[.4in]{,} \vec{\beta}\cdot \vec{\gamma}\neq 0 
\;.
\label{sca}
\end{equation}
We can expand $c=p+r$, $q_I = p_I + r_I$, where $p$, $p_I$ contain the Lie basis elements $T_p$ and $E_\sigma$, with $\vec{\beta} \cdot \vec{\sigma} = 0$, while $r$, $r_I$ contain  $E_\gamma$, with $\vec{\beta} \cdot \vec{\gamma} \neq 0$. Therefore, $q_I^{S_0} = p_I + r_I^{S_0}$, 
$  q_I^{S_0}|_w =p_I $. In addition, using the algebraic properties of the Cartan decomposition, it is easy to see that $[p_I,p] $ can only be a combination of the $E_\sigma$'s, so this term does not contribute to the scalar product in eq. (\ref{sca}). In this regard, note that $[p_I,p] $ contains
commutators of the form $[T_q, E_\sigma] \propto E_\sigma$, and $[E_\sigma, E_{\sigma'}] \propto E_{\sigma +\sigma'}$, which obviously satisfies $\vec{\beta} \cdot (\vec{\sigma}+\vec{\sigma}')=0$. As a consequence, BRST symmetry of the boundary conditions requires that the off-diagonal components of $c$, $c_I$, that are charged with respect to the weight $\vec{\beta}$, be also vanishing at the worldsheets. In turn, it can be verified that the induced boundary conditions are BRST invariant. Similar regularity conditions on the scalar fields must be imposed on those worldlines where $S_0$ leads to monopole-like singularities.

\section{Discussion and Conclusions}

In this work, we analyzed the BRST symmetry in different topological sectors of pure Yang-Mills theory. For this objective, motivated by lattice gauge fixing procedures designed to avoid the Gribov problem,
we introduced in the continuum a gauge fixing based on the classical solutions for a tuple of adjoint fields, with a flavour index, in the presence of the gauge field $\mathscr{A}_\mu$.

The action for this tuple possesses an $SU(N) \to Z(N)$ SSB pattern that strongly correlates $\mathscr{A}_\mu$ with a mapping $S\in SU(N)$, obtained from a generalized polar decomposition of the tuple. Next, the gauge fixing transforms $S$ to a given reference $S_0$, producing a partition of gauge field configurations into physically inequivalent sectors $\mathscr{V}(S_0)$, associated with different distributions of topological magnetic configurations. Then, we showed how to implement the gauge fixing of  $Z_{YM}^{(S_0)}$, the partial contributions to the partition function when path-integrating over the sector $\mathscr{V}(S_0)$. 
In particular, keeping the pure Yang-Mills dynamics requires not only the presence of ghost and antighost fields $c$, $\bar{c}$, but also flavoured versions $c_I$, $\bar{c}_I$, with their own transformation properties. 

The lesson to be learnt is that while there is a BRST symmetry that has the same functional form on each sector $\mathscr{V}(S_0)$, the class of ghost fields depends on the particular sector considered. This is a consequence of requiring the BRST invariance of the boundary conditions needed for the tuples of adjoint fields be well-defined everywhere. This induces
different boundary conditions on ghost fields that depend on the particular distribution of magnetic configurations such as closed center vortex worldsheets and their  magnetic weights.
 A similar result would apply to sectors formed by center vortices with different weights attached to interpolating monopoles. The expected consequences are twofold. On the one hand, the BRST symmetry on a given sector would be useful to analyze important questions such as renormalization and the independence of the partial contributions to the partition function on the gauge fixing parameters. On the other hand, the lack of a globally defined BRST symmetry is a nonstandard feature with consequences on the discussion of the physical states of the theory. The point is that the usual relation between BRST symmetry and quantum physical states rely on having a space of fields with a unique set of boundary conditions, defined from the beginning. 
Canonical quantization is usually implemented by expanding the field operators in field modes, which are used to construct the conserved BRST charge operator and analyze the asymptotic physical spectrum. For instance, this can be done in the case of noncompact QED with boundary conditions on a given distribution of conductors. In our case, if we were to ignore all  sectors $\mathscr{V}(S_0)$ with $S_0\in SU(N)$ containing defects, only considering the sector $\mathscr{V}(I)$ formed by gauge fields associated with regular mappings $S$, there would be no special boundary conditions, and the fields could be expanded in plane waves (or perturbed plane waves). In this perturbative sector, due to the quartet mechanism, the space of physical states would correspond to asymptotic transverse gluons, that is, no gluon confinement. 

In the complete theory, a classical conserved BRST charge could be obtained and evaluated for a given field configuration. However, at the quantum level, there is an ensemble of infinitely many sectors, with their own boundary conditions and expansion modes. How to relate the sector dependent BRST symmetry with the physical states of the theory is not at all clear. An important step towards this quest would be to gain information about the partial contributions and the definition of the ensemble integration.  In this respect, the integration over monopoles  with adjoint charges, and phenomenological dimensionful parameters,  was done in ref. \cite{GBO}. This reproduces some terms of a natural effective model with $SU(N) \to Z(N)$ SSB where gluons are seen as confined dual monopole configurations \cite{oxman}.

\section*{Acknowledgements}
We are grateful to Cesar Fosco, Rodrigo Sobreiro, and Silvio Sorella for useful discussions. The Conselho Nacional de Desenvolvimento Cient\'{\i}fico e Tecnol\'{o}gico (CNPq), the Coordena\c c\~ao de Aperfei\c coamento de Pessoal de N\'{\i}vel Superior (CAPES), and the Funda\c c\~{a}o de Amparo \`{a} Pesquisa do Estado do Rio de Janeiro (FAPERJ) are acknowledged for the financial support.

\end{document}